\documentclass[12pt,preprint]{aastex}
\usepackage{natbib}
\usepackage{graphicx}
\shorttitle{Yukawa Potential interacting dark matter}
\begin{document}
\title{Observational evidences of the Yukawa Potential Interacting Dark 
Matter}
\author{M.~H.~Chan}
\affil{Department of Physics and Institute of Theoretical Physics,
\\ The Chinese University of Hong Kong,
\\  Shatin, New Territories, Hong Kong, China}
\email{mhchan@phy.cuhk.edu.hk}
\begin{abstract}
Recent observations in galaxies and clusters indicate dark matter 
density profiles exhibit core-like structures which contradict to the 
numerical simulation results of collisionless cold dark matter. On the 
other hand, it has 
been shown that cold dark matter particles interacting through a Yukawa 
potential could naturally explain the cores in dwarf galaxies. In this 
article, I use the Yukawa Potential interacting dark matter model to 
derive two simple scaling relations on the galactic and cluster scales 
respectively, which give 
excellent agreements with observations. Also, in our model, the masses of 
the force carrier and dark matter particle can be constrained by the 
observational data. 
\end{abstract} 

\section{Introduction}
The nature of dark matter remains a fundamental problem in 
astrophysics and cosmology. The rotation curves of galaxies and the 
derived mass 
profiles in clusters indicate the existence of dark matter. It is commonly 
believed that dark matter is collisionless and becomes 
non-relativistic after decoupling. Therefore, they are regarded as cold 
dark matter (CDM). The CDM model can provide excellent fits on large scale 
structure observations such as Ly$\alpha$ spectrum \citep{Croft,Spergel}, 
2dF Galaxy 
Redshift Survey \citep{Peacock} and Cosmic Microwave Background 
\citep{Spergel2}. 

However, N-body simulations based on the CDM theory predict that 
the density profile of the collisionless dark matter halo should be 
singular at the center ($\rho \sim r^{-1}$) \citep{Navarro} while 
observations in dwarf galaxies give $\rho \sim r^{-0.29 \pm 0.07}$ 
\citep{Oh,Loeb}. On the cluster scale, observational data from 
gravitational lensing also show that cores exist 
in some clusters \citep{Tyson,Newman}. In particular, \citet{Sand} get 
$\alpha=-0.45 \pm 0.2$ by the combination of gravitational lensing and 
dynamical data of clusters MS2137-23 and Abell 383. Clearly, observations 
do not support the numerical small-scale predictions by the CDM model. 
This discrepancy is known as the core-cusp problem \citep{deBlok2}. 

Many theories have been invoked to solve the core-cusp problem. For 
example, some baryonic processes such as supernova feedback have been 
suggested to alleviate 
the problem \citep{Weinberg,deBlok2}. Recent high resolution cosmological 
hydro-dynamical simulations show that cored dark matter density profile 
could be produced in Milky-Way like halo if there is enough radiation 
pressure of massive stars before they explode as supernovae 
\citep{Maccio}. However, it is still controversial to make the conclusion 
because the actual contribution of supernova explosions is limited by the 
low star formation efficiency \citep{Penarrubia}. Also, it is challenging 
to invoke baryonic processes as the main mechanisms for some dark matter 
dominated galaxies \citep{Vogelsberger}. Another spectacular idea is 
that dark matter is not cold. The existence of keV sterile neutrinos, as a 
candidate 
of warm dark matter (WDM), has been proposed to solve the discrepancies 
\citep{Bode,Xue,Cho}. However, recent observations tend to reject the keV 
sterile neutrinos to 
be the major component of dark matter since the observational bound of 
sterile neutrino mass in Lyman-alpha forest contradicts to that in x-ray 
background \citep{Abazajian,Viel,Seljak}. Also, the WDM model alone cannot 
get a good 
agreement on the large scale power spectrum \citep{Spergel,Boyarsky}. 
Therefore, the success of the CDM 
model on large scales suggests that a modification of the dark matter 
properties may be the only approach to solve the discrepancies 
\citep{Spergel}. \citet{Spergel} proposed that the conflict of 
observations and simulations can be reconciled if the CDM particles are 
self-interacting. Later, \citet{Burkert,Yoshida} performed numerical 
simulations of self-interacting dark matter (SIDM) with constant cross 
section and showed that core-like structures could be produced. However, 
this proposal 
fell out of favour because gravitational lensing and X-ray data indicate 
that the cores of clusters are dense and ellipsoidal where SIDM model 
predicts that to be shallow and spherical \citep{Loeb}. Also, there is a 
discrepancy on the required cross section per unit mass in dwarf 
galaxies ($\sigma/m \sim 0.1-1$ cm$^2$ g$^{-1}$) and clusters ($\sigma/m 
\sim 0.01-0.1$ cm$^2$ g$^{-1}$) \citep{Miralda,Randall,Buckley,Tulin}. 
Recent numerical simulations indicate that only a small window open 
for a constant cross section SIDM model to work as a distinct alternative 
to CDM \citep{Zavala}. Therefore, a velocity-dependent cross section of 
the SIDM were explored to tackle the problems. 

\citet{Loeb} proposed that the possible existence of a Yukawa potential 
among the dark matter particles can resolve the problem raised by the SIDM 
with constant cross ssection. The velocity dependence could make 
scattering important in dwarf galaxies but unimportant in clusters 
\citep{Vogelsberger}. In this model, the dark matter particles of mass $m$ 
is set by an attractive Yukawa potential with coupling strength $\alpha$ 
mediated by a gauge boson of mass $m_{\phi}$ in the dark sector 
\citep{Feng,Loeb}. The cross section is well fitted by
\begin{equation}
\sigma \approx \left \{ \begin{array}{lll}
\frac{4 \pi}{m_{\phi}^2} \beta^2 \ln (1+ \beta^{-1}),       & {\ \ \beta 
\le 0.1,} \\ &\\
\frac{8 \pi}{m_{\phi}^2} \beta^2/(1+1.5 \beta^{1.65}),       & {\ \ 0.1 
\le \beta \le 10^3,} \\ &\\
\frac{\pi}{m_{\phi}^2}(\ln \beta +1-0.5 \ln^{-1} \beta)^2,       & {\ \ 
\beta \ge 10^3,} \end{array} \right.
\end{equation}
where $\beta=\pi v_{max}^2/v^2=2 \alpha m_{\phi}/(mv^2)$ and $v$ is the 
relative velocity of the dark matter particles \citep{Loeb}. The $v_{max}$ 
is the 
velocity at which $\sigma v$ reaches its maximum value. In the original 
proposal of the Yukawa interacting dark matter model, 
the parameter $v_{max}$ was considered to be $v_{max} \sim 10-100$ km/s in 
order to match the characteristic velocities of dwarf galaixes and 
clusters so that the scattering rate is larger in dwarf galaxies but 
much smaller in clusters. Nevertheless, this is not a necessary 
range. In Fig.~1, we can see that the cross section per unit mass is 
still larger on the galactic scale ($v \sim 10-100$ km/s) even if $v_{max} 
\sim 10^4$ km/s. Also, the scattering rate of dark matter particle is 
$\rho_c(\sigma/m)v$, which is still larger in galaxies because the central 
density $\rho_c$ is about 10 times larger than that in clusters. In this 
article, I release the free parameter $v_{max}$ to 
$\sim 10^4$ km/s so that the cross section will be $\sigma \propto (\ln 
\beta+1)^2$ and $\sigma \propto \beta^{-0.35}$ on the galactic 
($\beta \ge 10^3$) and cluster scales ($\beta \sim 10^2$) respectively. 
Therefore, the cross section is more velocity dependent on the cluster 
scale. In the following, I will use the above velocity-dependent 
self-interacting dark matter (vdSIDM) model with $v_{max} \sim 10^4$ 
km/s to derive a scaling relation to relate the total mass of dark matter 
halo $M$ and $v$. Then, I will compare the derived scaling relations with 
empirical fits on the galactic and cluster scales.

\section{vdSIDM model in galaxies and clusters}
In the vdSIDM model, the size of a core $R_c$ in a structure depends on 
the self-interacting rate of dark matter particles. Inside the core, 
we may assume that the dark matter particles interact with each other at 
least once during the evolution of a galaxy or cluster. Therefore, at 
$r=R_c$, we have
\begin{equation}
\rho_c \left( \frac{\sigma}{m} \right)v \approx H_0,
\end{equation}
where $\rho_c=3M_c/4\pi R_c^3$ is the central density of the core, $M_c$ 
is the total mass of the core and $H_0$ is the Hubble constant 
\citep{Rocha}. By using 
the Virial relation $v \approx \sqrt{GM_c/R_c}$ and Eq.~(2), we get
\begin{equation}
M_c=\left( \frac{3}{4\pi} \right)^{1/2} \left( \frac{\sigma}{m} 
\right)^{1/2}H_0^{-1/2}G^{-3/2}v^{3.5}.
\end{equation}
In galaxies, since the core mass is about one-tenth of the total mass 
($M_c \sim 0.1M$) \citep{Rocha}, we have
\begin{equation}
\log \left(\frac{M}{M_{\odot}} \right)=3.65+0.5 \log \left( 
\frac{\sigma/m}{0.1~\rm cm^2g^{-1}} \right)+3.5 \log \left( \frac{v}{\rm 
1~km/s} \right).
\end{equation}
For $\beta \ge 10^3$, the cross section in Eq.~(1) can also be 
approximated by using a powerlaw of $v$ (see Fig.~2):
\begin{equation}
\log \left( \frac{\sigma}{m} \right)=-0.37 \log \left( \frac{v}{1~\rm 
km~s^{-1}} \right)+ \log \left( \frac{\sigma_0}{m} \right),
\end{equation}
where $\sigma_0$ is a constant which depends on $m_{\phi}$. By putting the 
above equation into Eq.~(4), we get
\begin{equation}
\log \left(\frac{M}{M_{\odot}} \right)=3.65+0.5 \log \left( 
\frac{\sigma_0/m}{0.1~\rm cm^2g^{-1}} \right)+3.3 \log \left( \frac{v}{\rm 
1~km/s} \right).
\end{equation}   
On the other hand, the empirical fit of the baryonic Faber-Jackson 
relation obtained from a representative sample of 436 galaxies is given by 
$\log (v/{\rm 1~km~s^{-1}})=0.299 
\log(M_B/M_{\odot})-1.053$ \citep{Catinella}. Since $M_B \approx 0.17M$, 
the observed Faber-Jackson relation becomes $\log(M/M_{\odot})=4.29+3.34 
\log(v/1~\rm km~s^{-1})$. Compare the empirical fit with Eq.~(6), we 
get $\sigma_0/m \approx 1.9$ cm$^2$ g$^{-1}$. By Eq.~(5), the cross 
section 
per unit mass for dwarf galaxies ($v \sim 50$ km/s) and Milky-Way size 
galaxies ($v \sim 200$ km/s) are $\sigma/m \sim 0.4$ cm$^2$ g$^{-1}$ and 
$\sigma/m 
\sim 0.2$ cm$^2$ g$^{-1}$ respectively. Therefore, both the power-law 
dependence of the scaling relation ($\approx 3.3$) and the order of 
magnitude of the cross section per unit mass ($\sigma/m \sim 0.1-1$ 
cm$^2$ g$^{-1}$) are 
generally agree with the recent observations \citep{Buckley,Peter,Tulin}. 
Futhermore, since $\beta \ge 10^3$ in galaxies, by using $\sigma_0/m 
\approx 1.9$ cm$^2$ g$^{-1}$ and Eq.~(1), we have $m_{\phi}^2m \approx 
\pi(\ln \beta+1)^2/(\sigma/m) \sim 0.3$ 
GeV$^3$. By combining the above result with the estimated lower bound 
derived from dwarf galaxies $m_{\phi} >40$ MeV \citep{Buckley}, we have $m 
\le 200$ GeV. 

Similarly, we can apply the same model to clusters. However, 
since $v \sim 10^3$ km/s ($1 \ll \beta \le 10^3$) in clusters, by Eq.~(1), 
the cross section drops faster with velocity (see Fig.~1):
\begin{equation}
\frac{\sigma}{m}=\frac{8 \pi}{m_{\phi}^2m} 
\frac{\beta^2}{(1+1.5\beta^{1.65})} 
\approx \left(\frac{\sigma_0'}{m} \right) \left( \frac{v_{max}}{v} 
\right)^{0.7}. 
\end{equation}
By using the result $m_{\phi}^2m \sim 0.3$ GeV$^3$, we have 
$\sigma_0'/m=0.012$ cm$^2$ g$^{-1}$. Therefore, the 
cross section per unit mass in clusters is $\sigma/m \sim 0.06$ cm$^2$ 
g$^{-1}$, which is consistent with the recent observed bounds $\sigma/m 
\le 0.1$ cm$^2$ g$^{-1}$ \citep{Miralda,Randall,Buckley,Peter,Tulin}. 
Since the size of a cluster is about 100 times of the core size, which is 
equivalent to $M \sim 100M_c$ \citep{Arabadjis,Rocha}, by putting Eq.~(7) 
into Eq.~(3), we have
\begin{equation}
M=100 \left(\frac{3}{4 \pi} \right)^{1/2}H_0^{-1/2}G^{-3/2}v_{max}^{0.35} 
\left( \frac{\sigma_0}{m} \right)^{1/2}v^{3.15}.
\end{equation}
Since $v \approx \sqrt{3kT/m_g}$, where $T$ and $m_g$ are the temperature 
and mean mass of a hot gas particle, we can obtain a scaling relation  
\begin{equation}
M \approx 1.7 \times 10^{14}M_{\odot} \left( \frac{T}{2~\rm keV} 
\right)^{1.58}. 
\end{equation}
Surprisingly, this derived scaling relation gives excellent 
agreements with both the power dependence and proportionality 
constant of the empirical fits from 118 clusters $M=(1.56 \pm 0.01) 
\times 10^{14}M_{\odot}(T/2~\rm keV)^{1.57 \pm 0.06}$ \citep{Ventimiglia}. 

\section{Discussion}
The original purpose of suggesting the vdSIDM model is to explain the 
observed cores in dwarf galaxies without 
affecting the dynamics in clusters \citep{Loeb}. They assume $v_{max} 
\sim 10-100$ km/s so that the maximum cross section lies on the 
galactic scale \citep{Vogelsberger}. In fact, this is a free parameter 
which depends on $m_{\phi}$ and $m$, and it is not necessary to be about 
10-100 km/s. \citet{Rocha} show that $\rho_c \sim 
0.015M_{\odot}~{\rm pc^{-3}}(v/\rm 100~km~s^{-1})^{-0.55}$ by 
simulations, which means the central density of dark matter is higher in 
dwarf galaxies. As a result, the scattering rate of dark matter particle 
$\sim \rho_c(\sigma/m)v$ is always larger in dwarf galaxies than clusters 
even if $v_{max} \sim 10^4$ km/s. On the other hand, the 
circular velocity of a dwarf galaxy can be obtained by substituting 
Eq.~(5) into Eq.~(2), which gives $v \approx (4 \pi 
GH_0/3)^{0.38}R^{0.76}(\sigma_0/m)^{-0.38}$. For $R=250$ pc and $R=500$ 
pc, we get $v=13$ km/s and $v=22$ km/s respectively. These values are 
consistent with the observed circular velocities on the dwarf spheroidal 
scale \citep{Walker,Wolf}. Therefore, the large $v_{max}$ can still solve 
the too big to fail problem suggested by \citet{Boylan,Vogelsberger}.

In this article, I show that if $v_{max} 
\sim 10^4$ km/s, the cross section goes like $\sim v^{-0.37}$ and $\sim 
v^{-0.7}$ in galactic and cluster scales respectively. The derived scaling 
relation on the galactic scale 
is $M \propto v^{3.3}$, which agrees with observations $M \propto 
v^{3.34}$ \citep{Catinella}. The 
cross section per unit mass contrained by this model is $\sigma/m \sim 
0.2-0.4$ cm$^2$ g$^{-1}$ for $v=50-200$ km/s, which is also consistent 
with the observed bounds in dwarf and normal galaxies 
\citep{Buckley,Peter,Tulin}. By applying the same model in clusters, we 
get $M \propto T^{1.58}$, which again gives excellent agreements 
with observations in both proportionality constant ($\sim 
10^{14}M_{\odot}$) and power dependence ($1.57 \pm 0.06$) 
\citep{Ventimiglia}. These results provide evidences on the
non-power-law velocity dependent cross section of self-interacting dark 
matter. If $v_{max}$ is 10-100 km/s, the derived scaling relations would 
be $M \propto v^{3.15}$ and $M \propto T^{0.75}$ on the galactic and 
cluster scales respectively. Obviously, they do not match the empirical 
fits from observational data. Futhermore, in 
my model, it predicts $m_{\phi}^2m \sim 0.3$ GeV$^3$. If $m_{\phi} \ge 
40$ MeV, then $m \le 200$ GeV, which is a testable range in the 
future large hadron collision experiments. 

Recently, \citet{Vogelsberger2} study the impact of self-interacting 
dark matter on the velocity distribution of dark matter haloes and the 
anticipated direct detection signals. They find that all SIDM and vdSIDM 
models show departure from the velocity distribution of the CDM model in 
the center of the Milky Way halo. Therefore, different SIDM scenarios, 
including my model, might be distinguished from each other through the 
details of direct detection signals in the future \citep{Vogelsberger2}.

\section{Acknowledgements}
I am grateful to the referee for helpful comments on the manuscript.

\begin{figure*}
\vskip5mm
 \includegraphics[width=84mm]{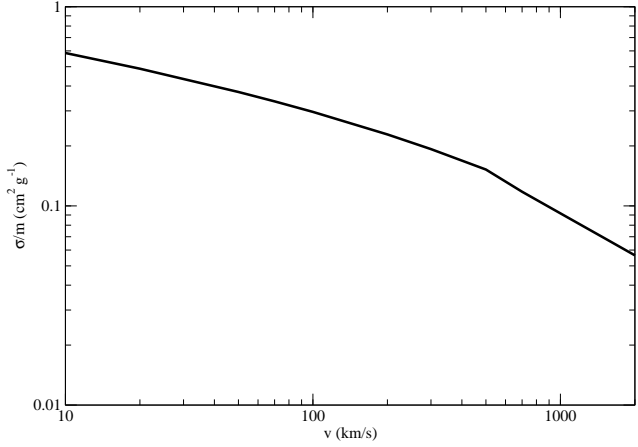}
 \caption{$\sigma/m$ versus $v$ for $v=50-1000$ km/s, $v_{max}=10^4$ 
km/s and $m_{\phi}^2m=0.3$ GeV$^3$.}
\end{figure*}

\begin{figure*}
\vskip5mm
 \includegraphics[width=84mm]{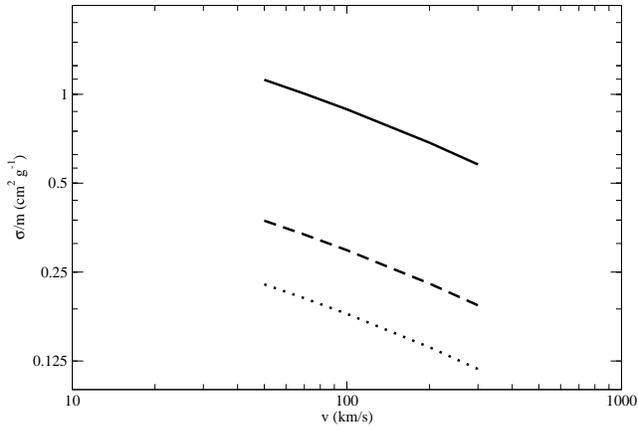}
 \caption{The solid line, dashed line and dotted line represent 
$\sigma/m$ versus $v$ for $m_{\phi}^2m=0.1$ GeV$^3$, 0.3 GeV$^3$ and 0.5 
GeV$^3$ respectively. The slopes of the lines are all $-0.37$.}
\end{figure*}

\end{document}